\documentclass[%
twocolumn,secnumarabic, nobibnotes, aps, prl, superscriptaddress,
 amsmath,amssymb
]{revtex4-2}

\usepackage{siunitx}

\usepackage{graphicx}% Include figure files
\usepackage{dcolumn}% Align table columns on decimal point
\usepackage{bm}% bold math
\begin{document}

%\preprint{APS/123-QED}

\title{Fast Gain Dynamics in Interband Cascade Lasers}

\author{Florian~\surname{Pilat}}
\email{florian.pilat@tuwien.ac.at}
%\affiliation{Institute of Solid State Electronics, TU Wien, Gusshausstrasse 25-25a, 1040 Vienna, Austria} 
\author{Andreas Windischhofer}
%\affiliation{Institute of Solid State Electronics, TU Wien, Gusshausstrasse 25-25a, 1040 Vienna, Austria}
\author{Maximilian Beiser}
\affiliation{Institute of Solid State Electronics, TU Wien, Gusshausstrasse 25-25a, 1040 Vienna, Austria}
\author{Vito F. Pecile}
\affiliation{Faculty of Physics, Faculty Center for Nano Structure Research, Christian Doppler Laboratory for Mid-IR Spectroscopy, University of Vienna, Boltzmanngasse 5, 1090 Vienna, Austria}
\affiliation{Vienna Doctoral School in Physics, University of Vienna, Boltzmanngasse 5, 1090 Vienna, Austria }
\author{Elizaveta Gangrskaia}
%\affiliation{Institute of  Photonics, TU Wien, Gusshausstrasse 27-29, 1040 Vienna, Austria}
\author{Audrius Pugžlys}
\affiliation{Institute of  Photonics, TU Wien, Gusshausstrasse 27-29, 1040 Vienna, Austria}
\author{Robert Weih}
%\affiliation{Nanoplus Nanosystems and Technologies GmbH, Oberer Kirschberg 4, 97218 Gerbrunn, Germany}
\author{Johannes Koeth}
\affiliation{Nanoplus Nanosystems and Technologies GmbH, Oberer Kirschberg 4, 97218 Gerbrunn, Germany}
\author{Andrius Baltuška}
\affiliation{Institute of  Photonics, TU Wien, Gusshausstrasse 27-29, 1040 Vienna, Austria}
\author{Oliver H. Heckl}
\affiliation{Faculty of Physics, Faculty Center for Nano Structure Research, Christian Doppler Laboratory for Mid-IR Spectroscopy, University of Vienna, Boltzmanngasse 5, 1090 Vienna, Austria}
\author{Benedikt Schwarz}
\email{benedikt.schwarz@tuwien.ac.at}
\affiliation{Institute of Solid State Electronics, TU Wien, Gusshausstrasse 25-25a, 1040 Vienna, Austria}
%\affiliation{John A. Paulson School of Engineering and Applied Sciences, Harvard University, Cambridge, Massachusetts 02138 USA}

\date{\today}

\begin{abstract}
Interband Cascade Lasers~(ICL) have matured into a versatile technological platform in the mid-infrared spectral domain. To broaden their applicability even further, ongoing research pursues the emission of ultrashort pulses. The most promising approach is passive mode-locking with a fast saturable absorber. However, despite vast attempts no passive mode-locked ICL has been demonstrated up to date. In this study we perform pump-probe measurements on the ICL gain and show that its dynamics are mainly ($\sim$70\%) governed by a fast recovery time of \qty[mode = text]{2}{\pico \second}. Our findings explain why passive mode-locking has not succeeded so far, shedding a new light on ICL dynamics as we propose strategies to overcome the current limitations.
\end{abstract}

\maketitle

%\section{Introduction}

The first experimental demonstration of the Quantum Cascade Laser~(QCL) in 1994 by Faist et al.~\cite{Faist1994} heralded the dawn of compact, electrically driven mid-infrared~(IR) lasers. An offspring of this revolutionary technology was proposed in the same year by Rui Yang~\cite{Yang1995}, which later became known as the Interband Cascade Laser. It combines the cascading nature of the QCL with an optical transition of a type-II diode laser. Charge carriers are generated at a semi-metallic interface~(SMI). Subsequently, the electrons and holes are transferred to the active transition quantum wells over their respective injector regions. Several important milestones mark the development of ICLs. Among these are the implementation of an active W-quantum well~(QW) transition~\cite{Felix1997}, a double QW hole injector~\cite{Olafsen1998} and carrier rebalancing by highly doping the electron injector wells~\cite{Vurgaftman2011}.
State-of-the-art ICLs are capable of continuous wave operation at room temperature~\cite{Kim2008} at a broad range of wavelengths in the mid-IR spectral domain from 3 to above \qty[mode = text]{6}{\micro \meter}~\cite{Nauschutz2023, Meyer2020}. This part of the electromagnetic spectrum is of special interest for high-precision spectroscopy~\cite{Vurgaftman2016,Sterczewski2017,Sterczewski2024}.

This field can benefit substantially from optical frequency comb sources. A broadband emission spectrum of multiple distinct and sharp longitudinal laser modes with stable phase relations between these modes enables elaborate heterodyne measurement schemes such as dual-comb spectroscopy. In the mid-IR spectral region, free-running Fabry-Pérot~(FP) QCLs~\cite{Hugi2012} and ICLs~\cite{Sterczewski2017, Schwarz2019} offer platforms for the generation of frequency combs. In contrast to one of the most prominent techniques for comb generation - passive mode-locking with a saturable absorber - these types of lasers achieve mode-locking through their intrinsic gain non-linearities~\cite{Opacak2021}. The resulting frequency combs show very different time dynamics compared to traditional passive mode-locking. Their intermodal phases are not zero, but span the whole range of $\mathrm{2 \pi}$~\cite{Schwarz2019}. This leads to a quasi-continuous emission - dubbed frequency-modulated~(FM) comb, rather than ultrashort pulses - also referred to as amplitude-modulated~(AM) combs.

While FM-combs are suited perfectly for methods such as linear absorption spectroscopy, some processes, e.g. non-linear spectral broadening and supercontinuum generation require high intensity pulses. Such an AM-comb source can be realized by fulfilling the two main conditions for passive mode-locking with a saturable absorber: Firstly, it requires the saturable absorber in the laser cavity to saturate stronger than the gain. Secondly, the absorber needs to have dynamics that are faster than the round-trip time of the cavity and the gain dynamics must be sufficiently slow in comparison~\cite{Haus1981}. The latter requirement disqualifies QCLs for this mode-locking technique, as their gain dynamics are in the picosecond range due to the fast intersubband transport, whereas the cavity roundtrip time is typically in the order of \qty[mode = text]{100}{\pico \second}. However, ICLs seemed to be perfect candidates as interband transitions are associated with longer lifetimes. Therefore, tremendous efforts have been made to passively mode-lock ICLs. However, frequency combs reported so far have shown FM-behaviour rather than short pulses~\cite{Meyer2020, Sterczewski2021}. Since no pulses could be observed, the absorber - an unbiased or negatively biased section made of the active material - was believed to be too slow. Still, making the absorber faster through ion bombardment~\cite{Bagheri2018} did not bear fruit.

In this work, we shed light on the absence of ICL pulses by answering the question whether the gain dynamics of ICLs are indeed sufficiently slow for passive mode-locking with a saturable absorber. Additionally, in the outlook, we propose strategies towards passively mode-locked ICLs.

\begin{figure}[t]
    \centering
    \includegraphics[width = 1.\columnwidth]{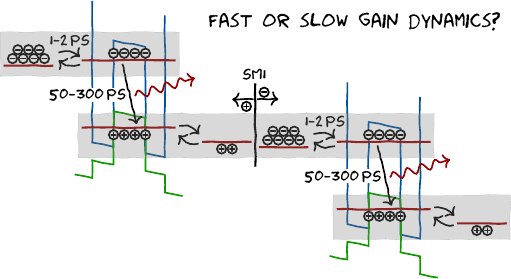}
    \caption{Sketch of two stages of the ICL band structure: conduction band~(blue) and valence band~(green) of the photo-active regions, with their corresponding electron and hole energy levels~(red). The black arrows indicate charge carrier transport, the red ones emitted photons. Charge carriers are generated at the semi-metallic interface~(SMI). Typical values for the electron lifetimes between injector and active region, as well as the interband transition lifetimes are indicated respectively. The question arises, whether the overall gain dynamics are fast or slow.}
    \label{fig_intro}
\end{figure}

In order to gain insight into the expected timescales, we start with the working principles of ICLs, sketched in Fig.~\ref{fig_intro}. Electron-hole pairs are generated at the SMI, which is realized by coupling hole and electron type subbands at a type-II broken gap semiconductor interface. Since it has been shown that high n-doping in the electron injector can significantly improve the performance of ICLs~\cite{Vurgaftman2011}, most state-of-the-art devices utilize this design approach. The higher electron population is indicated by a larger number compared to holes in the injectors. The injection of the generated electrons and holes into their respective laser levels takes place through fast intersubband scattering and tunneling. We want to emphasize here, that this intersubband scattering occurs in both directions, which leads to a very efficient balancing process between the injector and the laser level. The fast intersubband balancing regions are highlighted in Fig.~\ref{fig_intro} with grey rectangles. For electrons, this efficient process leads to lifetimes of picoseconds, in contrast to the much slower lifetimes of the interband transition of 50-\qty[mode = text]{300}{\pico \second} due to Auger scattering. The question arises: \textit{To what extent do these fast scattering processes contribute to the overall gain recovery dynamics, or are the dynamics still governed by the slow interband transition, as previously assumed?}

%\section{Methods}

\begin{figure*}[t]
    \centering
    \includegraphics[width = 2.\columnwidth]{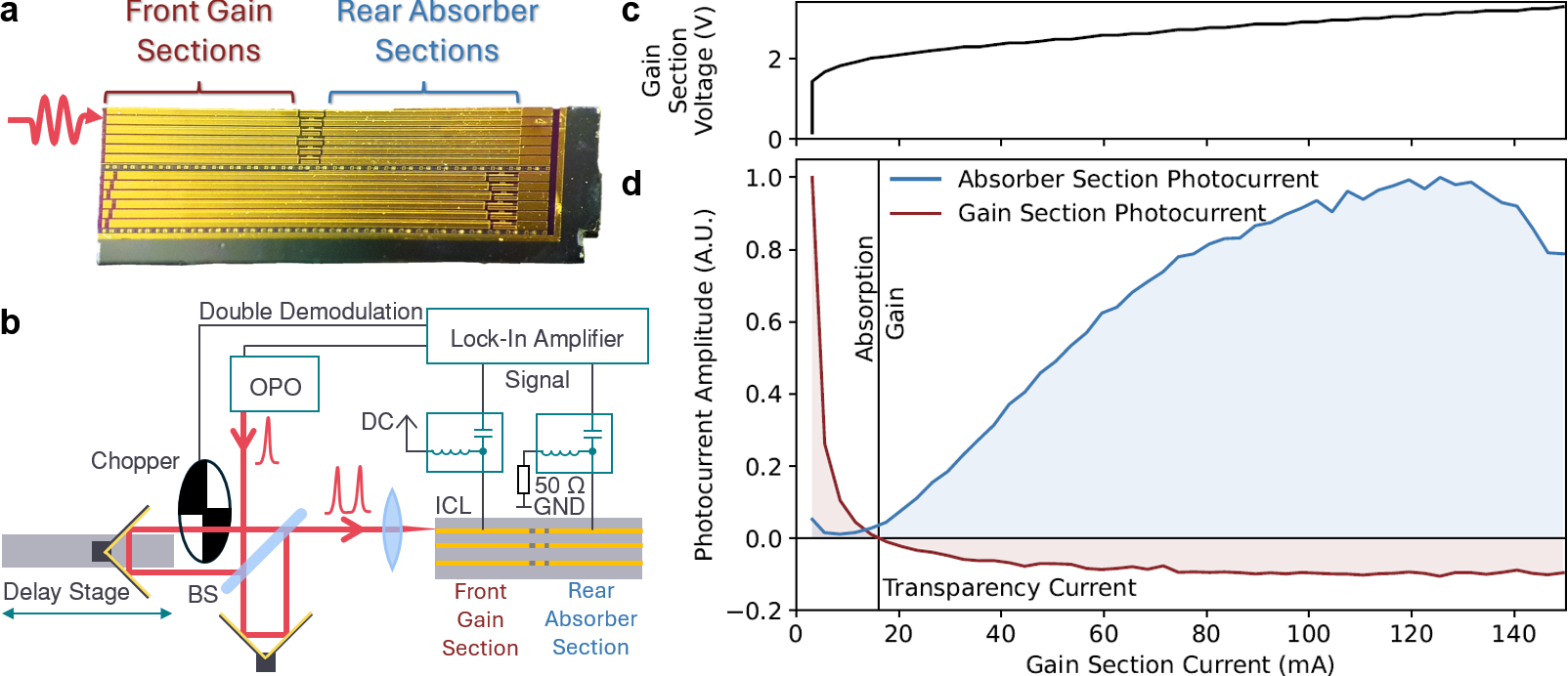}
    \caption{\textbf{a.} Microscope image of the ICL device under investigation. Several laser cavities are visible. An incident light pulse is illustrated as a red arrow, indicating the front facet of the measured device. It consists of a single ridge cavity with two electrically-separated top contacts: the front gain region and the rear absorber region.
    \textbf{b.} Pump-Probe Setup: Light pulses are generated from an optical parametric oscillator~(OPO), they are split and combined at a beam splitter~(BS), whereas one of them is sent to an optical delay stage, as well as a chopper. A lens is used to focus the combined beams onto the ICL facet. Bias-Ts are used to contact the front gain and the rear absorber sections. The gain section can be DC-biased with a laser driver, the absorber section is connected to ground with a \qty[mode = text]{50}{\ohm} resistor. The AC channels can be connected to a lock-in amplifier. The chopper frequency as well as the OPO trigger are used in a double demodulation scheme.
    \textbf{c.} ICL DC voltage over driving current characteristics.
    \textbf{d.} Experimental verification, that the gain is probed due to pulse amplification: AC-photocurrent responses to only probe pulses focused into the ICL cavity in dependence of the gain section DC bias current. The front gain section response~(red) gives the gain transparency current at \qty[mode = text]{16}{\milli \ampere}, at which no signal is measured. The incoming pulses experience absorption below the transparency current and gain above it. The blue line gives the absorber section response.}
    \label{fig_mehods}
\end{figure*}

In order to find an answer, we performed pump-probe experiments to investigate the time dynamics of the ICL gain.
For this purpose, we designed and fabricated special ICL devices that are optimized towards this challenging endeavor. 
The \qty[mode = text]{8}{\milli \meter} long and \qty[mode = text]{6}{\micro \meter} wide FP-ICL device used in this work is depicted in Fig.~\ref{fig_mehods}a. The red arrow indicates a light pulse impinging on the laser facet.
The laser consists of one long ridge cavity with separate contact sections.
When all sections are biased above threshold, the laser emits light around \qty[mode = text]{3.3}{\micro \meter}, so the gain also has to be probed at this wavelength. In the following experiments, however, the back section is not biased, but connected to ground via a \qty[mode = text]{50}{\ohm} resistor. This enables it to absorb the light sufficiently, so that no cavity effects i.e., interference from a back-reflected light wave, emerge. The device was thermoelectrically cooled to a temperature of \qty[mode = text]{15}{\celsius} and a DC bias was applied to the front section via a bias T, while measuring the pulse response over its AC port with a Zurich Instruments \mbox{UHFLI} lock-in amplifier. A ppqSense QubeCL was used as a low-noise current source and temperature controller for the ICL. The rear absorber section can also be connected to the lock-in using a bias-T, while its DC channel remains connected to the ground with a \qty[mode = text]{50}{\ohm} resistor. This is sketched in Fig.~\ref{fig_mehods}b along with the pump-probe setup. A custom built optical parametric oscillator~(OPO) system~\cite{Pecile2024} provided $\sim$\qty[mode = text]{500}{\femto \second} pulses at a wavelength \qty[mode = text]{3.3}{\micro \meter}, where the ICL has its gain peak. The pulses have a repetition rate of \qty[mode = text]{125.4}{\mega \hertz} and were set to an average power of \qty[mode = text]{100}{\milli \watt}. They were guided to a beam splitter, where one part is aligned to a Newport DL325 optical delay stage and a chopper. The other part has a fixed delay and both are combined again at a beam splitter. Subsequently, they are focused onto the ICL waveguide facet using a high-numerical aperture aspheric lens~(Thorlabs AL72512-E1). The whole optical path was kept as short as possible to mitigate the influence of beam divergence. The lock-in amplifier uses the OPO pulse trigger and the chopper signal for a double demodulation scheme.

\begin{figure*}[t]
    \centering
    \includegraphics[width = 2.\columnwidth]{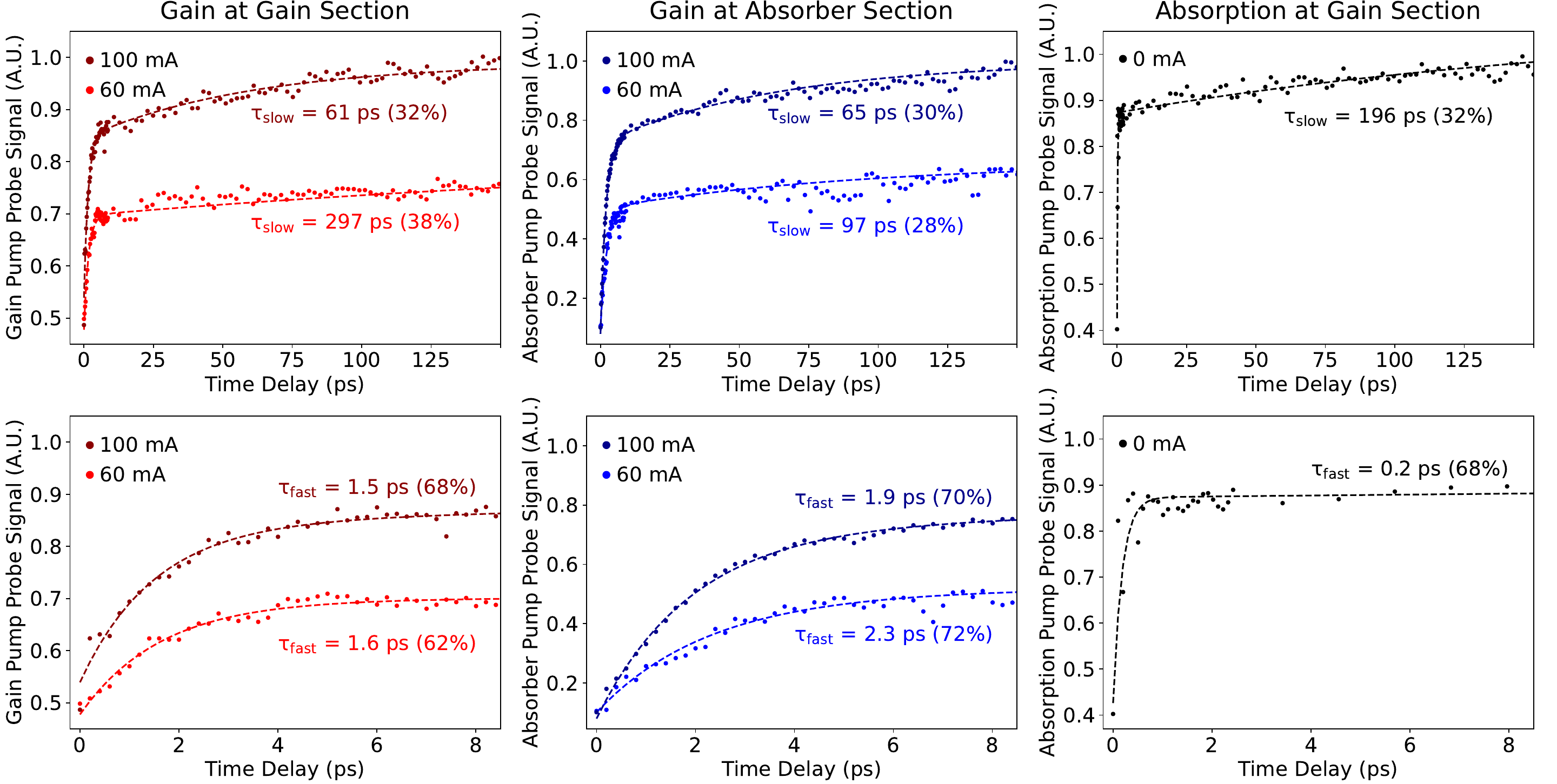}
    \caption{Results of the ICL pump-probe experiments: 
     Red plots show the gain section response at 60 and \qty[mode = text]{100}{\milli \ampere} DC driving current, blue plots show the absorber section response for these driving currents. They correspond to the gain dynamics. The black plots show the unbiased gain section response, corresponding to the absorption dynamics. The points indicate the measured data, the dashed line is a double exponential fit. The fitted time constants together with their respective percentage weight are written next to the curves.
    The top row shows the range of \qty[mode = text]{150}{\pico \second}, the bottom row the first \qty[mode = text]{8}{\pico \second}. The gain recovery is dominated ($\sim$70\%) by a \qty[mode = text]{2}{\pico \second} fast component.}
    \label{fig_results}
\end{figure*}

%\section{Results and Discussion}

Upfront it has to be mentioned that the measurements pose quite some challenges. One has to make absolutely sure to indeed measure the gain dynamics rather than other effects, like substrate absorption or metal-semiconductor interfaces at the contacts. Due to the tiny laser facet with dimensions of merely $\sim$\qty[mode = text]{5}{\micro \meter}, a highly stable source, a high-numerical aperture lens and careful alignment is required. The pulse power has to be high enough to reach sufficient saturation and signal-to-noise ratio, but low enough not to damage the laser structure. The pulse repetition rate of the source has to be much higher compared to drift mechanisms of the whole system to employ reasonable averaging.
Additionally, a careful alignment procedure was performed before each measurement to account for long-term drifts.
While blocking the stationary arm of the interferometer and moving the delay stage to the zero path difference position, the front facet of the laser was aligned using the chopped beam. The signal at the rear section was maximized, while positively biasing the front section to ensure that the light couples into the waveguide and experiences gain.

In order to confirm that indeed the gain is probed, a bias current sweep was performed and the pulse response photocurrent signals at the front and rear section were measured. Fig.~\ref{fig_mehods}c shows the DC voltage over the DC bias current of the front section, which has a typical shape for an ICL.
In Fig.~\ref{fig_mehods}d the corresponding AC photocurrent signals of the front gain~(red) and the rear absorber~(blue) section are plotted. At zero bias, the front section acts as a photodetector. It absorbs all the light and generates a photocurrent. With an increasing DC bias current, the signal gets weaker as charge carriers are brought into the active region and fewer electron-hole pairs are generated via photon absorption. At a current of \qty[mode = text]{16}{\milli \ampere} the signal reaches zero and undergoes the expected sign flip. Here, the gain medium is on the verge of inversion, i.e. transparent, and the light just experiences the waveguide losses. At higher DC bias current the impinging pulses get amplified due to stimulated emission, which causes charge carriers to flow in the opposite direction compared to the detector operation. A saturation effect can be seen for higher bias currents.
The absorber section acts as a photodetector all the time. It shows almost no signal at zero bias, since most of the pulse is absorbed in the front section, as indicated in Fig.~\ref{fig_mehods}d in blue. At the gain transparency current already some signal can be observed. For higher bias currents the pulse experiences gain in the front section and an increasing signal can be observed. This amplified detector scheme has been investigated in more detail by Dal Cin et al. ~\cite{DalCin2024}. Here, a similar geometry is used, however, with a longer detector section to sufficiently remove cavity effects. Again, the signal saturates at higher currents until it decreases again. Since the laser was aligned at \qty[mode = text]{100}{\milli \ampere} for this measurement, at other bias currents some temperature-induced drift might have occurred, causing a misalignment. This might be the reason for the small signal at zero bias, or the decreasing signal at high bias currents. Additionally, at high bias currents the absorber section might not attenuate the strong pulse sufficiently, allowing for cavity effects.

With the confirmation that the pulses experience gain, the pump-probe experiments were performed according to the sketch in Fig.~\ref{fig_mehods}b. The photocurrent response was measured both at the front section as well as the rear section. The front signal corresponds to the charge carriers flowing into the laser after stimulated emission caused by the pulses, whereas the rear signal is the photocurrent generated from absorption of the amplified pulses. As the pulses deplete the charge carriers in the gain section, less signal is detected, when both arrive at the same time, i.e. at zero path difference.
The farther the pulses are apart in time, the more the gain recovers, leading to an increase in signal. The time constants that can be observed this way correspond to the gain dynamics.

The results for both sections are depicted in Fig.~\ref{fig_results} for 60 and \qty[mode = text]{100}{\milli \ampere} DC bias current at the gain section. The top row shows data up to a time delay of \qty[mode = text]{150}{\pico \second}, the bottom row shows the first \qty[mode = text]{8}{\pico \second}, to visualize the fast features. In order to account for the slightly diverging beam and possible minor misalignment of the setup, which can cause an overall decrease of the signal with increasing time delay, a linear fit has been made to a data subset at long time delays, where the influence due to the gain dynamics is negligible. Subsequently, this fit was subtracted from the whole data set. 
The obtained curves are displayed as dots in Fig.~\ref{fig_results} and can be adequately fitted with a double exponential function~(dashed lines). The fitting time constants $\mathrm{\tau_{slow}}$ and $\mathrm{\tau_{fast}}$ and their corresponding contributions for front and rear sections are shown in the graphs. In all four cases there is a very fast component of around \qty[mode = text]{2}{\pico \second}, which makes up roughly 70\% of the overall signal. This is a clear indication that a strong portion of the gain dynamics is governed by the fast intersubband transport, similar to QCLs.
The slow parts are in the order of 60 to \qty[mode = text]{100}{\pico \second}, with an exception for the gain section data at \qty[mode = text]{60}{\milli \ampere}, which is almost \qty[mode = text]{300}{\pico \second}. We attribute that effect to fluctuations in the setup causing measurement artifacts, as it was optimized for short delays. Since the beam is slightly divergent and the alignment is very sensitive to small deviations, the setup is not ideal to measure long time delays with high precision. Therefore, the values for the long contributions should be considered with higher uncertainties.
There is a trend, that higher currents lead to overall faster dynamics. This can be attributed to Auger scattering, which increases with higher populations.
Additionally, it seems, that the signal at the absorber section is slightly slower. This might be due to additional non-linear effects in the absorber section. Higher overall intensity, which is present further away from zero path delay might saturate the absorber resulting in longer appearing time constants.

The absorption dynamics at the gain section - the pump-probe photocurrent for \qty[mode = text]{0}{\milli \ampere} DC bias current - were investigated. They are depicted in the right two plots of Fig.~\ref{fig_results} in black. Here, also two time constants can be fitted, with comparable magnitudes. 
Here, the fast component is significantly shorter with \qty[mode = text]{200}{\femto \second}, which could be partially caused by the autocorrelation signals present due to the short pulse duration of the OPO. However, as this signal is not appearing on both sides of the baseline, this effect is not yet fully understood and motivates a more detailed future investigation with potentially shorter pulses.
Summarizing, the presented measurements provide strong evidence regarding the presence of extremely fast carrier dynamics in the absorbing regime, which can be explained by the efficient intersubband transport.

%The fast component, however, is significantly shorter . Since the pulses are $\sim$\qty[mode = text]{500}{\femto \second}, the observed signal might partly contain the interferometric autocorrelation signal. There, however, a signal should be present at both sides of the baseline. Nevertheless, the results hint towards the presence of extremely fast carrier dynamics in the absorbing regime, which can be explained by the efficient intersubband transport, but should be investigated further in future works.

%\section{Conclusion and Outlook}

In conclusion, we performed pump-probe experiments on an ICL and our findings show, that about 70\% of the gain dynamics are governed by a fast recovery of \qty[mode = text]{2}{\pico \second}, similar to QCLs. We consistently show, that we indeed probe the gain of our device employing a specialized cavity design and a dedicated alignment routine. The dynamics are measured twofold, the impact of impinging pulses is measured at the laser medium, where the population dynamics directly reflect the gain dynamics. Additionally, the amplified pulses are measured after leaving the active gain medium, comparable to more traditional pump-probe techniques. Both results are in good agreement.
Our findings offer a deeper understanding of the dynamics governing ICLs and clearly show, why passive mode-locking has not been achieved so far. 

On the contrary, a promising conclusion can be drawn from our results. Novel types of frequency combs found in ring-cavity QCLs~\cite{Piccardo2019,Meng2021}, especially the Quantum Walk Laser~\cite{Heckelmann2023} and Nozaki-Bekki solitons~\cite{Opacak2024} heavily rely on fast gain dynamics. The latter benefits from a high linewidth enhancement factor, which would make ICLs the ideal platform for these combs. As the fast gain component is linked to carrier injection, rather than the laser transition itself, we suspect that energy can be stored in the ICL gain medium. This suggests that a monolithic master oscillator power amplifier~(MOPA) scheme for soliton amplification can be efficiently realized with ICLs, ultimately leading to higher pulse power levels than currently possible with QCLs.
Additionally, the fast dynamics are of interest for high-speed applications, such as free-space optical communications~\cite{Didier2023}.

One question remains - can the ICL band structure be redesigned to suppress the fast gain component, enabling passive mode-locking? The study of Feng et al.~\cite{Feng2018} shows passive mode-locking with similar lasing transitions that lack the fast injector. We suspect the pool of electrons in the injector due to the carrier rebalancing~\cite{Vurgaftman2011} combined with the efficient intersubband transport into the upper laser level to play a crucial role in the fast ICL dynamics. This suggests that a dedicated slow ICL design might bear fruit, opening up a completely new path in the pursuit of passive mode-locked ICLs.

%overcoming the current deadlock and paving a new way towards passive mode-locked ICLs.

\section*{Funding} This work was supported by the European Research Council~(ERC) project MonoComb~[853014] under the European Union’s Horizon 2020 research and innovation programme, by the European Innovation Council~(EIC) Pathfinder project UNISON~[101128598] under the European Union’s Horizon Europe programme and by the Austrian Science Fund~(FWF) [DOI: 10.55776/P36040]. The financial support by the Austrian Federal Ministry for Digital and Economic Affairs, the National Foundation for Research, Technology and Development and the Christian Doppler Research Association is gratefully acknowledged.

\section*{Disclosures} The authors declare no conflicts of interest.

\section*{Data Availability} Data underlying the results presented in this paper may be obtained from the authors upon reasonable request.

\bibliography{main}

\end{document}